\documentclass[preprint,showpacs,showkeys,preprintnumbers,amsmath,amssymb,superscriptaddress]{revtex4}
\usepackage{graphicx}
\usepackage{dcolumn}
\usepackage{bm}

\begin{document}
\title{Some examples of different descriptions of energy-momentum density
in the context of Bianchi IX cosmological model}

\author{Z. Nourinezhad}
\email{z.nourinejad@bandaranzaliiau.ac.ir}

\affiliation{Islamic Azad University, Bandaranzali Branch,
Bandaranzali, Iran}

\author{S. Hamid Mehdipour}

\email{mehdipour@liau.ac.ir}

\affiliation{Islamic Azad University, Lahijan Branch, P. O. Box
1616, Lahijan, Iran}

\date{\today}

\begin{abstract}
Based on the Bianchi type IX metric, we calculate the energy and
momentum density components of the gravitational field for the five
different definitions of energy-momentum, namely, Tolman,
Papapetrou, Landau-Lifshitz, M{\o}ller and Weinberg. The energy
densities of M{\o}ller and Weinberg become zero for the spacetime
under consideration. In the other prescriptions, i.e., Tolman,
Papapetrou and Landau-Lifshitz complexes, we find different
non-vanishing energy-momentum densities for the given spacetime,
supporting the well-known argument in General Relativity that the
different definitions may lead to different distributions even in
the same spacetime background.
\end{abstract}


\maketitle

\section{\label{sec:1}Introduction}
The interpretation of energy-momentum as a significant conserved
quantity is one of the most interesting and stimulating problems in
the theory of Einstein General Relativity (GR). There have been
numerous efforts to acquire a well-defined illustration for the
energy and momentum localization in the literature. Unfortunately,
there is still no prevalent accepted interpretation of energy and
momentum distributions in GR \cite{Xul}.

The energy-momentum conservation in GR can be written as
\begin{equation}
\label{mat:1} \nabla _{\mu}T^ {\mu}_ {\nu} = 0,    \quad (\mu,
\nu=0, 1, 2, 3),
\end{equation}
where $T^{\mu}_{\nu}$ indicates the symmetric energy-momentum tensor
including the matter and all non-gravitational fields. In 1915,
Einstein \cite{Ein} acquired an expression for the energy-momentum
complexes comprised of the contribution from gravitational field
energy by introducing the energy-momentum $t_{\nu}^{\mu}$ which is
not a tensor and is called the gravitational field pseudotensor. The
energy-momentum complex satisfies the local conservation laws, i.e.
\begin{equation}
\label{mat:2}{\cal{T}}^{\mu}_{\nu \,\, , \mu}\equiv
\frac{\partial}{\partial x^{\mu}} \left(\sqrt{-g} (T_{\nu}^{\mu}
+t_{\nu}^{\mu})\right) =0,
\end{equation}
where the energy-momentum tensor $T^{\mu}_{\nu}$ is replaced by the
energy-momentum complex ${\cal{T}}^{\mu}_{\nu}$ which is a
combination of the tensor $T^{\mu}_{\nu}$ plus the pseudotensor
$t_{\nu}^{\mu}$ but in the ordinary form of conservation laws. So,
we have
\begin{equation}
\label{mat:3}{\cal{T}}^{\mu}_{\nu}=\theta^{\mu\lambda}_{\nu \,\,\,\,
, \lambda},
\end{equation}
where $\theta^{\mu\lambda}_{\nu}$ are denoted as the superpotentials
and are not uniquely determined.

With an appropriate choice of a coordinates system, the pseudotensor
$t_{\nu}^{\mu}$ can be identified such to disappear at a special
point. Schrodinger demonstrated that the pseudotensor can vanish
outside the schwarzschild radius utilizing an appropriate choice of
coordinates. There have been many efforts in order to attain a more
fitting quantity to illustrate the distribution of energy-momentum
on account of matter, non-gravitational fields and gravitational
field pseudotensor. Einstein supported the expression of
pseudotensor to portray the gravitational field and explained that
this energy-momentum pseudo-complex prepares reasonable expressions
for the complete energy-momentum of the closed systems. Many authors
have prescribed different explanations for the energy-momentum
complex \cite{Ein,Tol,Pap,Lan,Ber,Gol,Mol,Wei}. These explanations
can only give significant outcomes if the computations are carried
out in cartesian coordinates. In 1982, Penrose \cite{Pen} initiated
the proposal of quasi-local energy to find the energy-momentum of a
curved spacetime by using any coordinate system. Many theoretical
physicists \cite{Vir} considered an assortment of different
proposals of the quasi-local energy to study different models of the
universe. Very general results for the most general nonstatic
spherically symmetric metric is known by Virbhadra \cite{Vir3}.
However these proposals of the energy-momentum complexes could not
lead to some unique definition of energy in GR because each of these
quasi-local expressions have their own issues.

In this paper, we present some well-known energy-momentum densities
based on Bianchi IX cosmological model to study the problem of
localization of the energy and momentum in GR. The spatially
homogeneous and anisotropic Bianchi models play a significant role
in modern cosmology. However, on account of intricate essence of the
field equations, there are minor works on anisotropic models (see,
e.g., \cite{Mish} and also \cite{Wam}), particularly on the Bianchi
IX cosmological model \cite{Bel} (the so-called Mixmaster universe
\cite{Mis}). Several generalizations to the Mixmaster universe have
also been considered in detail, by some authors (for a comprehensive
review, see \cite{Dam}). The Bianchi universes provide useful
paradigms to investigate the nonlinear behavior of the Einstein
equations, due to the time-dependency of the gravitational fields
(see, for example, \cite{Pra} and references therein). Studying on
the anisotropic models was first considered after finding the
anisotropic behavior of the microwave background radiation. In other
words, no decisive confirmation is perceived that the early universe
had essentially the same properties on the early era. Hence, it is
credibly applicable to study the universes established upon the
different Bianchi type metrics. A general study of the dynamical
properties of anisotropic Bianchi universes in the context of GR is
presented by Perez \cite{Per}. In this setup, the Bianchi type VIII
and IX universes are dynamically equivalent. In 2005, the authors in
Ref.~\cite{Khe} applied the canonical quantum theory of gravity
(Quantum Geometrodynamics) to the homogeneous Bianchi type IX
cosmological model, and accordingly they developed the framework for
the quantum theory of homogeneous cosmology. In 2009, Bakas {\it et
al} \cite{Bak} considered spatially homogeneous (but generally
non-isotropic) cosmologies in the Ho\v{r}ava-Lifshitz gravity and
compared them to those of GR using Hamiltonian methods. They
exhibited that the Mixmaster dynamics is completely dominated by the
quadratic Cotton tensor potential term for very small volume of the
universe, by focusing on the closed-space cosmological model
(Bianchi type IX). Recently, Damour and Spindel \cite{Dam2} have
studied the mini–superspace quantization of spatially homogeneous
(Bianchi) cosmological universes sourced by a Dirac spinor field.
They presented the exact quantum solution of the Bianchi type II
system and discussed the main qualitative features of the quantum
dynamics of the (classically chaotic) Bianchi type IX system. Barrow
and Yamamoto \cite{Bar}, in a recent paper, have investigated the
stability of the Einstein static universe as a non-LRS Bianchi type
IX solution of the Einstein equations in the presence of both
non-tilted and tilted fluids. They have found that the static
universe is unstable due to homogeneous perturbations of Bianchi
type IX to the future and the past.

Moreover, the comprehension of type IX solutions, due to its
anisotropy, with their oscillatory treatment in the direction of the
initial singularity is prevalently related to one of the keys
towards a more clear comprehension of singularities in GR and thus
could be an interesting candidate to test the quantum theory
\cite{Bel2}. We therefore focus on Bianchi IX cosmological model
which demonstrates a specifically rich dynamical structure. The
Bianchi type IX universe is defined by the line element:
\begin{equation}
\label{mat:4} ds^2 = -dt^2 + S^2(t)dx^2 + R^2(t)\left[dy^2 +\sin^2y~
dz^2\right]-S^2(t)\cos y\left[2dx-\cos y~ dz\right]dz,
\end{equation}
where the functions $S$ and $R$ are function in $t$ and determined
from the field equations. We will use the above line element to
acquire the energy and momentum densities in the next section.

The paper is organized in the following. In Section \ref{sec:2}, by
applying the energy-momentum definitions of Tolman, Papapetrou,
Landau-Lifshitz, M{\o}ller and Weinberg, we calculate the
energy-momentum densities of the universe based on Bianchi type
IX metric, respectively. Conclusion is presented in Section \ref{sec:3}.\\

\section{\label{sec:2}Energy-Momentum Complexes: some examples}
The energy-momentum in Tolman's prescription \cite{Tol} has the form
{\footnote{Throughout this paper, Latin indices ($i$, $j$, $\ldots$)
represent the spatial coordinate values while Greek indices ($\mu$,
$\nu$, $\ldots$) represent the spacetime labels. We set the
fundamental constants equal to unity; $G = c = 1$.}}
\begin{equation}
\label{mat:5}\Upsilon^\nu_\mu =
\frac{1}{8\pi}U^{\nu\lambda}_{\mu~\,\, ,\lambda},
\end{equation}
where the Tolman's superpotential $U^{\nu\lambda}_{\mu}$ is defined
by
\begin{equation}
\label{mat:6}U^{\nu\lambda}_{\mu}
=\sqrt{-g}\left(-g^{\kappa\nu}V^\lambda
_{\mu\kappa}+\frac{1}{2}g^{\nu}_\mu
g^{\kappa\vartheta}V^\lambda_{\kappa\vartheta}\right),
\end{equation}
with
\begin{equation}
\label{mat:7}V^\alpha_{\beta\gamma}=-\Gamma^\alpha_{\beta\gamma}+\frac{1}{2}g^{\alpha}_\beta\Gamma^\delta_{\delta\gamma}
+\frac{1}{2}g^\alpha_\gamma\Gamma^\delta_{\delta\beta}.
\end{equation}
The locally conserved energy-momentum complex of Tolman includes
contributions from the matter plus the all gravitational and
non-gravitational fields {\footnote{Note that the elements of the
energy-momentum complex containing matter plus all gravitational and
non-gravitational fields are the same in all other prescriptions as
follows in this paper.}}. The Tolman energy and momentum complex
satisfies the local conservation laws as follows,
\begin{equation}
\label{mat:8} \Upsilon^\nu_{\mu\, , \nu} = 0,
\end{equation}
where $\Upsilon^0_\mu$ is a combination of the energy-momentum
tensor including the matter and all non-gravitational fields plus
the gravitational field pseudotensor. Therefore we can describe the
quantity $\Upsilon^0_0$ as representing the energy density of the
whole physical system including gravitation, and describe the
quantity $\Upsilon^0_i$ as representing the components of the total
momentum density. In order to calculate the energy and momentum
density components for the Bianchi type IX metric, we need to
compute the essential non-zero components of $U^{0\lambda}_{\mu}$
which give
\begin{equation}\label{mat:9}
\begin{array}{l}
U^{02}_{0} = -S{(t)}\cos y, \\
U^{01}_{1} = \frac{1}{2}R{(t)}\sin y\left(\dot{S}(t) R{(t)}-2S{(t)} \dot{R}(t)\right), \\
U^{00}_{2} = -\frac{1}{2}S{(t)} R^{2}{(t)}\cos y, \\
U^{02}_{2} = U^{03}_{3} = -\frac{1}{2}R^{2}{(t)} \dot{S}(t)\sin y,\\
U^{01}_{3} = -R{(t)} \sin y  \cos y \left(\dot{S}(t) R{(t)}-S{(t)}\dot{R}(t)\right),\\
\end{array}
\end{equation}
where overdot abbreviates $\partial/\partial t$. Substituting these
values for Eq.~(\ref{mat:5}), we obtain the components of energy and
momentum density in the prescription of Tolman as follows
\begin{equation}
\label{mat:10} \Upsilon^{0}_{0}= \frac{S{(t)}}{8\pi},
\end{equation}
\begin{equation}
\label{mat:11} \Upsilon^{0}_{1} = \Upsilon^{0}_{3}=0, \,\,
\Upsilon^{0}_{2} = -\frac{R{(t)}\cos y}{8\pi}\left(\dot{S}(t)
R{(t)}+S(t)\dot{R}(t)\right).
\end{equation}

The energy and momentum in the prescription of Papapetrou \cite{Pap}
takes the form
\begin{equation}
\label{mat:13}\Omega^{\mu\nu} = \frac{1}{16\pi}
N^{\mu\nu\lambda\kappa} _{\,~~\quad,\lambda\kappa},
\end{equation}
where
\begin{equation}
\label{mat:14}N^{\mu\nu\lambda\kappa} =
\sqrt{-g}\left(g^{\mu\nu}\eta^{\lambda\kappa} -
g^{\mu\lambda}\eta^{\nu\kappa} + g^{\lambda\kappa}\eta^{\mu\nu} -
g^{\nu\kappa}\eta^{\mu\lambda}\right).
\end{equation}
The Papapetrou's superpotential $ N^{\mu\nu\lambda\kappa}$ is
symmetric on its first pair of indices with $\eta^{\mu\nu}$ that is
the Minkowski metric. The Papapetrou energy-momentum complex obeys
the local conservation laws,
\begin{equation}
\label{mat:15}\Omega^{\mu\nu}_{~~\, , \nu} = 0,
\end{equation}
where $\Omega^{00}$ and $\Omega^{i0}$ represent the energy and
momentum density components respectively. In this prescription, the
essential non-zero components of $N^{\mu0\lambda\kappa}$
corresponding to the metric (\ref{mat:4}) yield the following
expressions:
\begin{equation}\label{mat:16}
\begin{array}{l}
N^{0000} = 2S{(t)}R^{2}{(t)}\sin y,\\
N^{0011} = \frac{R^2(t)}{S(t)}\Big(S^2(t)-1\Big)\sin y-S(t)\cos y \cot y,\\
N^{1010} = -\frac{R^2(t)}{S(t)}\Big(S^2(t)+1\Big)\sin y-S(t)\cos y \cot y,\\
N^{3010} =N^{1030}= N^{0031} = N^{0013}=-S{(t)} \cot y, \\
N^{0022} = S(t) \big(R^2{(t)}-1\big)\sin y,  \\
N^{2020}= -S(t) \big(R^2{(t)}+1\big)\sin y, \\
N^{3030} = -\frac{S{(t)}}{\sin y}\Big(R^2{(t)}\sin^2 y +1 \Big), \\
N^{0033} = \frac{S{(t)}}{\sin y}\Big(R^2{(t)}\sin^2 y -1\Big). \\
\end{array}
\end{equation}
Replacing the above expressions in Eq.~(\ref{mat:13}), one can
obtain the energy and momentum densities in Papapetrou's
prescription. So, we have
\begin{equation}
\label{mat:17}\Omega^{00} = \frac{\pi \sin
y}{16}\left[8R{(t)}\dot{R}(t)\dot{S}(t)+2R^2{(t)}\ddot{S}(t)+S{(t)}\left(4\dot{R}^2(t)
+4R{(t)}\ddot{R}(t)-R^2{(t)}+1\right)\right],
\end{equation}
\begin{equation}
\label{mat:18}\Omega^{20} = \Omega^{30} = 0, \,\, \Omega^{10} =
-\frac{\pi \cos
y}{16}\left(\dot{S}(t)+R^2{(t)}\dot{S}(t)+2S{(t)}R{(t)}\dot{R}(t)\right).
\end{equation}

The energy and momentum in the prescription of Landau and Lifshitz
\cite{Lan} is given by
\begin{equation}
\label{mat:20} L^{\mu\nu} = \frac{1}{16\pi}{\cal
S}^{\mu\nu\lambda\kappa} _ {\,~~\quad ,\lambda\kappa}
\end{equation}
with
\begin{equation}
\label{mat:21}{\cal S}^{\mu\nu\lambda\kappa} =
-g\left(g^{\mu\nu}g^{\lambda\kappa} -
g^{\mu\lambda}g^{\nu\kappa}\right),
\end{equation}
where $L^{\mu\nu}$ is symmetric with respect to its indices and the
Landau-Lifshitz's superpotential ${\cal S}^{\mu\nu\lambda\kappa}$
has the symmetries similar to the curvature tensor. The
Landau-Lifshitz's energy-momentum complex $ L^{\mu\nu} $ confirms
the local conservation laws
\begin{equation}
\label{mat:22} L^{\mu\nu}_{~~\, , \nu} = 0.
\end{equation}
The $L^{00}$ is the energy density and $L^{0i}$ are the momentum
density components. The required non-vanishing components of ${\cal
S}^{0\nu\lambda\kappa}$, associated with the Bianchi type IX metric,
lead to the following relations:
\begin{equation}\label{mat:23}
\begin{array}{l}
{\cal S}^{0011} = -{\cal S}^{0101} = -R^2{(t)}\left(R^2{(t)}\sin^2 y+S^2{(t)}\cos^2 y\right), \\
{\cal S}^{0022} = -{\cal S}^{0202} = \sin^2 y \,\, {\cal S}^{0033} = -\sin^2 y \,\, {\cal S}^{0303} = -S^2{(t)}R^2{(t)}\sin^2 y,\\
{\cal S}^{0031} = {\cal S}^{0013} = -{\cal S}^{0103} = -{\cal S}^{0301}=-S^2{(t)} R^2{(t)}\cos y.\\
\end{array}
\end{equation}
Substituting these components for (\ref{mat:20}), one can obtain the
energy and momentum densities which give the following relations:
\begin{equation}
\label{mat:24} L^{00} = 2S^2{(t)}R^2{(t)}\left(\sin^2 y-\cos^2
y\right),
\end{equation}
\begin{equation}
\label{mat:25}  L^{01} = L^{03} =0, \,\, L^{02} = 2\sin
\left(2y\right)\left(R^2{(t)}S{(t)} \dot{S}(t)+ R{(t)}\dot{R}(t)
S^2{(t)}\right).
\end{equation}

The energy and momentum in the prescription of M{\o}ller \cite{Mol}
is given by
\begin{equation}
\label{mat:27}M^{\mu}_\nu=\frac{1}{8\pi}\chi^{\mu\lambda}_{\nu
\,\,\,\, ,\lambda },
\end{equation}
where the antisymmetric superpotential $ \chi^{\mu\lambda}_\nu $ has
the form
\begin{equation}
\label{mat:28}\chi^{\mu \lambda}_\nu=-\chi^{ \lambda\mu}_\nu=\sqrt{-
g}\left(g_{\nu \sigma, \kappa}- g_{\nu\kappa,
\sigma}\right)g^{\mu\kappa} g^{\lambda\sigma},
\end{equation}
where $g$ is the determinant of the metric $g_{\mu\nu}$. It can be
simply shown that the M{\o}ller's energy-momentum complex satisfies
the local conservation laws
\begin{equation}
\label{mat:29} M^\mu_{\nu\, , \mu} = 0.
\end{equation}
$M_0^0$ is the energy density and $M^0_i$ are the momentum density
components. For the line element given by Eq.~(\ref{mat:4}) the
required non-vanishing components of $\chi^{0\lambda}_{\nu}$ are
\begin{equation}\label{mat:30}
\begin{array}{l}
\chi^{01}_{1} = -2R^{2}{(t)} \dot{S}(t)\sin y,\\
\chi^{02}_{2}= \chi^{03}_{3}= -2S{(t)} R{(t)} \dot{R}(t)\sin y, \\
\chi^{01}_{3} = -R{(t)} \left(\dot{S}(t)R{(t)}-S{(t)} \dot{R}(t)\right)\sin (2y). \\
\end{array}
\end{equation}
Entering the above components in Eq.~(\ref{mat:27}), we can find the
energy and momentum densities as follows
\begin{equation}
\label{mat:31} M^{0}_{0} = 0,
\end{equation}
\begin{equation}
\label{mat:32} M^{0}_{1}=M^{0}_{3}=0, \,\, M^{0}_{2} =
-\frac{S{(t)}R{(t)}\dot{R}(t)\cos y}{4\pi}.
\end{equation}
The energy density is zero in M{\o}ller's prescription, as can be
seen from Eq.~(\ref{mat:31}).

The Weinberg's energy-momentum complex \cite{Wei} is expressed by
the equation
\begin{equation}
\label{mat:33}W^{\mu\nu} =\frac{1}{16\pi}\Delta^{\mu\nu\lambda}_
{~\quad, \lambda},
\end{equation}
where Weinberg's superpotential $\Delta^{\mu\nu\lambda}$ is
antisymmetric on its first pair of indices which defines as
\begin{equation}
\label{mat:34}\Delta^{\mu\nu\lambda} = \partial^\mu h^\kappa_\kappa
\eta^{\nu\lambda} - \partial^\nu h^\kappa_\kappa \eta^{\mu\lambda} -
\partial_\kappa h^{\kappa\mu} \eta^{\nu\lambda}
+ \partial_\kappa h^{\kappa\nu} \eta^{\mu\lambda} + \partial^\nu
h^{\mu\lambda} - \partial^\mu h^{\nu\lambda},
\end{equation}
where $\partial_\mu\equiv\partial/\partial x^\mu$,
$\partial^\mu\equiv\partial/\partial x_\mu$ and $h_{\mu\nu}$ shows
the symmetric tensor defined as
$h_{\mu\nu}=g_{\mu\nu}-\eta_{\mu\nu}$. The energy-momentum in
Weinberg's prescription satisfies the local conservation laws
\begin{equation}
\label{mat:35}W^{\mu\nu}_{~~\, , \nu} = 0.
\end{equation}
$W^{00}$ and $W^{i0}$ are the energy and momentum density components
respectively. The following non-zero components of
$\Delta^{\mu0\lambda}$ are required to find energy-momentum
densities in this prescription
\begin{equation}\label{mat:36}
\begin{array}{l}
\Delta^{101} = \frac{2}{R^7{(t)} S^7{(t)}\sin^6 y} \Big( R^4{(t)}
S^3{(t)} \dot{R}(t)\left(S^2(t)\cos^2 y+2\right)\sin^4 y\\
\qquad\quad+R^5{(t)} S^2{(t)}\dot{S}(t)\left(S^2(t)+2\right)\sin^4
y\cos^2 y+4S^5{(t)} R^2{(t)}\dot{R}(t) \sin^2 y\cos^2 y\\
\qquad\quad+R^7{(t)}\dot{S}(t)\left(S^2(t)+1\right)\sin^6 y+R^3{(t)}
S^4{(t)} \dot{S}(t)\sin^2 y \cos^2 y\left(\cos^2 y+1\right)\\
\qquad\quad +R^2{(t)} S^7{(t)} \dot{R}(t)\sin^2 y \cos^2
y\left(\cos^2 y+1\right)\\
\qquad\quad+S^7{(t)}\dot{R}(t) \cos^2 y \left(3\cos^2 y+\sin^2 y+1\right)\Big), \\
\Delta^{103} =\frac{\sin (2y)}{(R{(t)}S{(t)}\sin y)^7
}\Big(2R^2{(t)}S^3(t)\dot{R}(t)\sin^2 y+S^5(t)\dot{R}(t)\left(3\cos^2 y+\sin^2 y+1\right)\\
\qquad\quad+R^5(t)\dot{S}(t)\left(S^2(t)+1\right)\sin^4 y+R^3(t)S^2(t)\dot{S}(t)\sin^2 y \left(\cos^2 y+1\right)\\
\qquad\quad+R^2(t)S^5{(t)} \dot{R}(t)\sin^2 y \left(\cos^2
y+1\right)\Big),\\
\Delta^{200} = -\frac{4\cot y}{R^4{(t)}\sin^2 y}, \\
\Delta^{202} = \frac{2}{R^7{(t)} S^3{(t)}\sin^2 y}\Big(\dot{R}(t)
S^3{(t)} \cos^2 y +\dot{S}(t) R^3{(t)} \sin^2 y+\dot{R}(t)
R^3(t)\sin^2 y\\ \qquad\quad+\dot{R}(t) S^3{(t)}+R^2{(t)} S^3{(t)} \dot{R}(t)\sin^2 y\Big), \\
\Delta^{301}=\frac{ \cos y}{\left(S(t)\sin y\right)^4
}\left(R^2{(t)} \sin^2 y+S^2{(t)} \cos^2 y\right)
\Delta^{202}, \\
\Delta^{303}= \frac{1+\cos^2 y}{ \sin^4 y }\Delta^{202}. \\
\end{array}
\end{equation}
We find the components of energy and momentum density distribution
in the prescription of Weinberg as follows
\begin{equation}
\label{mat:37}W^{00} = 0,
\end{equation}
\begin{equation}
\label{mat:38}W^{10} = W^{30}=0, \,\, W^{20} =\frac{\pi \cot y
\dot{R}(t)\left(2R^2{(t)}-1\right)}{2R^7{(t)}\sin^2 y}.
\end{equation}
As can be seen from Eq.~(\ref{mat:37}), the energy density is zero
in Weinberg's prescription.

\section{\label{sec:3} Conclusion}
In conclusion, we have demonstrated some examples of the different
descriptions of the energy-momentum density in the context of
Bianchi IX cosmological model. We have found that the
energy-momentum complexes of M{\o}ller and Weinberg provide the zero
energy density in the gravitational background under consideration.
Also, it is possible to vanish both M{\o}ller's and Weinberg's
momentum density components at a specific spacetime point, e.g., $y
= \pi/2$. However, this vanishing by no means holds in general. So,
these results may be supported by this statement which tells us that
different energy and momentum complexes can give the same results
for the same gravitational background \cite{Vir2}. In the remaining
prescriptions, i.e., Tolman, Papapetrou and Landau-Lifshitz
complexes, we have acquired different non-zero energy and momentum
densities which sustain this statement that different energy and
momentum complexes could yield different energy and momentum
distributions for a given gravitational background \cite{Bergq}. In
fact, in all the prescriptions for the spacetime under
consideration, each of the different expressions might indicate a
physically and geometrically consequence connected to the boundary
conditions. Furthermore, according to the equivalence principle, the
appearance of pseudotensors as noncovariant objects is the origin of
these discrepancies which reflects the fact that the gravitational
field cannot be perceived at a point. Therefore, according to
Ref.~\cite{Coo}, the main outcome of this paper points out that
energy cannot be localized in this type of
time-dependent gravitational background of the spacetime.\\

\section*{Acknowledgments}
ZN thanks H. Farajollahi for discussions. We thank the anonymous
referees for their useful comments.

\end{document}